\documentclass[
aps,%
11pt,%
final,
notitlepage,%
oneside,%
twocolumn,%
nobibnotes,%
nofootinbib,%
preprintnumbers,%
superscriptaddress,%
showpacs,%
centertags,%
showkeys,%
amsmath,%
amssymb]{revtex4}

\usepackage{graphicx}
  \graphicspath{{../figures/}}
\usepackage{amsmath}

\newcommand{\m}[1]{\mathrm {#1}}
\newcommand{\unit}[1]{\;\m{#1}}

\begin{document}

\title{Parameter differences of the charged and neutral rho-meson family}

\author{E. Barto\v s}
\affiliation{Inst. Phys., Slovak Acad. Sci., D\'ubravsk\'a cesta 9, 845\,11 Bratislava, Slovak Republic}

\author{S. Dubni\v cka}
\affiliation{Inst. Phys., Slovak Acad. Sci., D\'ubravsk\'a cesta 9, 845\,11 Bratislava, Slovak Republic}

\author{A.-Z. Dubni\v ckov\'a}
\affiliation{FMPI Comenius University, Mlynsk\'a dolina, 842\,48 Bratislava, Slovak Republic}

\author{M. Fujikawa}
\affiliation{Nara Women's University, Kita uoya nishi machi, 630-8506 Nara, Japan}

\author{H. Hayashii}
\affiliation{Nara Women's University, Kita uoya nishi machi, 630-8506 Nara, Japan}


\begin{abstract}
New very precise KLOE data on $\m{e}^+\m{e}^-\to\pi^+\pi^-$ obtained by radiative return method in Frascati are unified with corrected CMD-2 and SND Novosibirsk $\m{e}^+\m{e}^-\to\pi^+\pi^-$ data and supplemented below and beyond by older data in order to be described by the unitary and analytic pion electromagnetic form factor model, which provides the most precise neutral $\rho$-meson family parameters.

Then the recently appeared accurate Belle (KEK) data of the weak pion form factor are described by the same unitary and analytic model, as it follows from the CVC hypothesis, providing the charged $\rho$-meson family parameters.
As a result the most reliable parameter differences of the $\rho$-meson family are determined.
\vspace{1pc}
\end{abstract}

\maketitle

\section{Introduction and motivation}\label{sec:intro}

Isotopic spin is in a very good approximation the conserved quantum number in strong interactions. A breaking of the corresponding symmetry occurs as a consequence of the electromagnetic (EM) interactions and the mass difference of the up and down quarks. Practically, it is demonstrated in nature by a splitting of hadrons into isomultiplets.

In this contribution we are concerned with $\rho$-meson resonances. The cleanest determination of their parameters comes from the $\m{e}^+\m{e}^-$ annihilation and $\tau$-lepton decay. However, as it is declared by the Review of Particle Physics \cite{Amsler:2008zzb}, experimental accuracy is not yet sufficient for unambiguous conclusions. The difference for $\rho^0$ and $\rho^\pm$ are presented in averaged to be $m_{\rho^0}-m_{\rho^\pm}=-0.7\pm0.8 \unit{MeV}$ and $\Gamma_{\rho^0}-\Gamma_{\rho^\pm}=0.3\pm1.3 \unit{MeV}$, respectively.

In our opinion the latter is changed in two aspects. On one hand, new very accurate KLOE data \cite{Aloisio:2004bu} on the pion EM form factor (FF) at the energy range $0.35\unit{GeV}^2\leq t\leq 0.95\unit{GeV}^2$ were obtained by the radiative return method in Frascati. Also corrected CMD-2 \cite{Akhmetshin:2006bx} and SND \cite{Achasov:2006vp} Novosibirsk $\m{e}^+\m{e}^-\to\pi^+\pi^-$ data appeared recently. On the other hand the weak pion FF accurate data \cite{Fujikawa:2008ma} from the measurement of the $\tau^-\to\pi^-\pi^0\nu_{\tau}$ decay by Belle (KEK) experiment were published. The unitary and analytic model \cite{Dubnicka:1995zk} of the pion EM FF, to be represented by one analytic function for $-\infty<t<+\infty$ was elaborated, which is always successfully applied for a description of existing data on the pion EM FF from $\m{e}^+\m{e}^-\to\pi^+\pi^-$ and due to the CVC hypothesis \cite{Dubnickova:1992js} equally well also for a description of existing data on the weak pion FF from $\tau^-\to\pi^-\pi^0\nu_{\tau}$ decay. As a result more sophisticated evaluation of difference of $\rho$-meson family parameters can be achieved.

\section{Unitary and analytic pion FF model} \label{sec:uam}

There is no theory able to predict the pion FF behavior up to now. Though on the role of a dynamical theory of strong interactions QCD is pretending, as a consequence of its asymptotic freedom, it gives just the asymptotic behavior
\begin{equation} \label{eq:as}
\m{F}_\pi(t)|_{t\to-\infty}\sim -\frac{16\pi f_\pi^2 \alpha_s(t)}{t},
\end{equation}
with the weak pion decay constant $f_\pi=92.4\pm0.2\unit{MeV}$ and QCD running coupling constant $\alpha_s(t)$.

Therefore a phenomenological approach, based on the synthesis of the experimental fact of a creation of $\rho$-meson family in $\m{e}^+\m{e}^-$ annihilation into two pions, the asymptotic behavior (\ref{eq:as}) and the analytic properties, is still the most successful way of a reconstruction o the pion EM FF behavior in the whole region of its definition.

This approach gives the expression (see \cite{Dubnicka:1995zk})
\begin{align} \label{eq:mod}
F_\pi[&W(t)]= \left( \frac{1-W^2}{1-W_N^2} \right)^2 \frac{(W-W_Z)(W_N-W_P)}{(W_N-W_Z)(W-W_P)} \nonumber\\
\times&\Big\{ \frac{(W_N-W_\rho)(W_N-W_\rho^*)}{(W-W_\rho)(W_-W_\rho^*)}\nonumber\\
\times&\frac{(W_N-1/W_\rho)(W_N-1/W_\rho^*)}{(W-1/W_\rho)(W-1/W_\rho^*)} \Big(\frac{f_{\rho\pi\pi}}{f_\rho}\Big)\nonumber\\
+&\sum\limits_{v=\rho',\rho''}\Big[\frac{(W_N-W_v)(W_N-W_v^*)}{(W-W_v)(W_-W_v^*)}\nonumber\\
\times&\frac{(W_N+W_v)(W_N+W_v^*)}{(W+W_v)(W+W_v^*)}\Big] \Big(\frac{f_{v\pi\pi}}{f_v}\Big)\Big\}
\end{align}
with the conformal mapping
\begin{align}
W(t)&= \unit{i}\frac{\sqrt{q_{in}+q}-\sqrt{q_{in}-q}}{\sqrt{q_{in}+q}+\sqrt{q_{in}-q}};\\ q&=\sqrt{\frac{t-t_0}{4}};\quad q_{in}=\sqrt{\frac{t_{in}-t_0}{4}} \nonumber
\end{align}
of four-sheeted Riemann surface into one $W$-plane, $W_Z$ and $W_P$ the zero and the pole, by means of which a contribution of the left-hend cut on the II. Riemann sheet is simulated, $f_{v\pi\pi}$ and $f_{v}$ the vector-meson-pion and the universal vector-meson coupling constants, respectively, whereby
\begin{align*}
\Big(\frac{f_{\rho'\pi\pi}}{f_\rho'}\Big) &= \frac{1}{\frac{N_{\rho'}}{|W_{\rho'}|^4}-\frac{N_{\rho''}}{|W_{\rho''}|^4}}\\
\times&\Big\{1-\Big(\frac{N_{\rho'}}{|W_{\rho'}|^4}-\Big[1+2\frac{W_Z W_P}{W_Z-W_P}\\
\times&\mathrm{Re}W_\rho(1-|W_\rho|^{-2})\Big]N_\rho\Big) \Big(\frac{f_{\rho\pi\pi}}{f_\rho}\Big)\Big\},\\
\Big(\frac{f_{\rho''\pi\pi}}{f_\rho''}\Big) &= \frac{1}{\frac{N_{\rho'}}{|W_{\rho'}|^4}-\frac{N_{\rho''}}{|W_{\rho''}|^4}}\\
\times&\Big\{-1+\Big(\frac{N_{\rho''}}{|W_{\rho''}|^4}-\Big[1+2\frac{W_Z W_P}{W_Z-W_P}\\
\times&\mathrm{Re}W_\rho(1-|W_\rho|^{-2})\Big]N_\rho\Big) \Big(\frac{f_{\rho\pi\pi}}{f_\rho}\Big)\Big\}
\end{align*}
and
\begin{align*}
N_\rho=& (W_N-W_\rho)(W_N-W_\rho^*)\\
&\times(W_N-1/W_\rho)(W_N-1/W_\rho^*),\\
N_v=& (W_N-W_v)(W_N-W_v^*)\\
&\times(W_N+W_v)(W_N+W_v^*);\quad v=\rho',\rho''.
\end{align*}

The model is defined on four-sheeted Riemann surface with complex conjugate poles (corresponding to unstable $\rho$-resonances) on unphysical sheets, and reflecting all known properties o the pion EM FF. It depends on 10 physically interpretable free parameters, $t_{in}$, $m_{\rho}$, $\Gamma_{\rho}$, $f_{\rho\pi\pi}/f_{\rho}$, $m_{\rho'}$, $\Gamma_{\rho'}$, $m_{\rho''}$, $\Gamma_{\rho''}$, $W_Z$ and $W_P$.

\section{Description of EM pion FF data}

The pion EM FF can be measured everywhere on the real axis of the $t$-plane and as a result there is almost continuous interval o 381 experimental points for $-9.77\unit{GeV}^2\leq t\leq 13.48\unit{GeV}^2$, which all are described by the pion EM FF model (\ref{eq:mod}) in the space-like and the time-like regions simultaneously. The most important from them are accurate KLOE data \cite{Aloisio:2004bu} at the energy range $0.35\unit{GeV}^2\leq t\leq 0.95\unit{GeV}^2$ obtained in Frascati by the radiative return method and also the corrected Novosibirsk CMD-2 data \cite{Akhmetshin:2006bx} at the range $0.36\unit{GeV}^2\leq t\leq 0.9409\unit{GeV}^2$ and SND data \cite{Achasov:2006vp} at the range $0.1521\unit{GeV}^2\leq t\leq 0.9409\unit{GeV}^2$, which can influence the finite relults substantially. They are supplemented at the interval $-9.77\unit{GeV}^2\leq t\leq 0.3364\unit{GeV}^2$ and $0.9557\unit{GeV}^2\leq t\leq 13.48\unit{GeV}^2$ by other existing data.

An application of the pion EM FF unitary an analytic model (\ref{eq:mod}) to the best description of all existing 381 experimental points was achieved and the values of the parameters are presented in Tab.~\ref{table:1}.

\section{Description of weak pion FF data}
\begin{table*}[tb]
\caption{The values of the fitting parameters for the fit of pion FF. The values are shown for two cases, the result of fitting $\m{e}^+\m{e}^-$ data to the U\&A model of pion EM FF (second column), the result of fitting $\tau^-$ data to the U\&A model of weak pion FF (third column). The differences of the values for both cases are presented in fourth column.}
\label{table:1}
\renewcommand{\tabcolsep}{0.5pc} 
\renewcommand{\arraystretch}{1.2} 
\centering
\begin{tabular}{@{}cccc}
\\[2pt]
\hline
Parameter & $\rho^0$ & $\rho^\pm$ & $\Delta$ ($\rho^\pm-\rho^0$) \\
\hline
$t_{in}$  & $(1.3646\pm 0.0198) \unit{GeV}^2$ & $(1.2432\pm 0.0157) \unit{GeV}^2$  & \\%
$W_Z$     & $0.1857\pm 0.0004$           & $0.4078\pm 0.0013$            & \\%
$W_P$     & $0.2335\pm 0.0005$           & $0.6197\pm 0.0007$            & \\%
$m_{\rho}$  & $(758.2260\pm 0.4620) \unit{MeV}$ & $(761.6000\pm 0.9520) \unit{MeV}$  &  $(3.3740\pm 1.0582) \unit{MeV}$ \\%
$m_{\rho'}$  & $(1342.3060\pm 46.6200) \unit{MeV}$ & $(1373.8340\pm 11.3680) \unit{MeV}$  &  $(31.5280\pm 47.9860) \unit{MeV}$ \\%
$m_{\rho''}$  & $(1718.5000\pm 65.4360) \unit{MeV}$ & $(1766.8000\pm 52.3600) \unit{MeV}$  &  $(48.3000\pm 83.8060) \unit{MeV}$ \\%
$\Gamma_{\rho}$  & $(144.5640\pm 0.7980) \unit{MeV}$ & $(139.9020\pm 0.4620) \unit{MeV}$  &  $(-4.6620\pm 0.8502) \unit{MeV}$ \\%
$\Gamma_{\rho'}$  & $(492.1700\pm 138.3760) \unit{MeV}$ & $(340.8720\pm 23.8420) \unit{MeV}$  &  $(-151.2980\pm 140.4150) \unit{MeV}$ \\%
$\Gamma_{\rho''}$  & $(489.5800\pm 16.9540) \unit{MeV}$ & $(414.7080\pm 119.4760) \unit{MeV}$  & $(-74.8720\pm 120.6729) \unit{MeV}$ \\%
$f_{\rho\pi\pi}/f_{\rho}$  & $1.0009\pm 0.0001$ & $0.9998\pm 0.0002$  & \\%
$\chi^2/\m{NDF}$  & 1.78 & 1.96  & \\%
\hline
\end{tabular}\\[2pt]
\end{table*}

Hadronic decays of the $\tau$-lepton provide a clean environment for studying the dynamics of hadronic states with various quantum numbers. Among them is $\tau^-\to\pi^-\pi^0\nu_{\tau}$, which is dominated by the $\rho^-$-meson family resonances and thus it can be used to extract information on the charged $\rho$-meson family properties.

From the conservation of vector current (CVC) theorem, the $\pi^-\pi^0$ mass spectrum in the $\tau^-\to\pi^-\pi^0\nu_{\tau}$ decay can be related to the total cross section of the $\m{e}^+\m{e}^-\to\pi^+\pi^-$ process, which is directly related with the pion EM FF. As a result the same unitary and analytic pion EM FF model can be applied to a description of the weak pion FF data, which can be drawn from the measured normalized invariant mass-spectrum.

Though there are measurements of $\tau^-\to\pi^-\pi^0\nu_{\tau}$ decay to be carried out previously by ALEPH \cite{Schael:2005am} and CLEO \cite{Anderson:1999ui}, in this contribution we are concentrated only on the high-statistics measurement \cite{Fujikawa:2008ma} of the weak pion FF from $\tau^-\to\pi^-\pi^0\nu_{\tau}$ decay with the Belle detector at the KEK-B asymmetric-energy $\m{e}^+\m{e}^-$ collider as they are charged by the lowest total errors.

An application of the pion EM FF unitary an analytic model (\ref{eq:mod}) to the best description of 62 experimental points \cite{Fujikawa:2008ma} on the weak pion FF was achieved by the values of the parameters as presented in Tab.~\ref{table:1}.

\section{Results and conclusions}\label{sec:res}

There may be a question about the fact that the obtained $\rho$-meson family parameters are differing from those presented at the Review of the Particle Physics \cite{Amsler:2008zzb}, especially of the $\rho$- and $\rho$'-mesons. As it is well known, the resonance parameters depend on the parametrization used in a fit of data. However, in this contribution it plays no crucial role as finally we are interested in a difference of the corresponding parameters. Moreover, in a determination of the parameters we exploit the unitary and analytic model of the pion FF, in which any resonance is defined as a pole on the unphysical sheets of the Riemann surface to be considered as the most sophisticated approach in a description of resonant states.

The difference of the $\rho$-meson family resonance parameters is presented in Tab.~\ref{table:1}. On the base of these results one can declare that the masses of the neutral $\rho$-meson family are lower than the masses of the charged $\rho$-meson family and the widths are just reversed. Considering the evaluated errors, one can confidently affirm only in the case of the $\rho(770)$-meson parameters and also in the case of the $\rho'(1450)$-meson widths, that for the charged and the neutral states they are different. For other $\rho$-meson family parameters one can say nothing definitely and one has to wait for even more precise experimental data.

\section*{Acknowledgments}

Some of the authors (E.B., S.D. and A.Z.D.) thank Slovak Grant Agency for Sciences VEGA for support under Grant No. 2/7116/29.

%

\end{document}